\documentclass{an}
\usepackage{graphicx}
\usepackage{times}
\usepackage{fancyhdr}
\begin{document}

\title{Biermann Mechanism in Primordial Supernova Remnant and Seed Magnetic Fields}

\author{Hidekazu Hanayama$^{1,2}$, Keitaro Takahashi$^{3}$,
Kei Kotake$^{4}$, Masamune Oguri$^{5,6}$,
Kiyotomo Ichiki$^{2}$, and Hiroshi Ohno$^{7}$}
\institute{
Department of Astronomy, School of Science,
University of Tokyo, Hongo 7-3-1, Bunkyo, Tokyo 113-0033, Japan.
\and
National Astronomical Observatory of Japan, Mitaka,
Tokyo 181-8588, Japan.
\and
Department of Physics, Princeton University, Princeton,
NJ 08544.
\and
Science and Engineering, Waseda University, 3-4-1 Okubo,
Shinjuku, Tokyo 169-8555, Japan.
\and
Department of Astrophysical Sciences, Princeton
University, Peyton Hall, Ivy Lane, Princeton, NJ 08544.
\and
Department of Physics, School of Science, University of Tokyo,
7-3-1 Hongo, Bunkyo, Tokyo 113-0033, Japan.
\and
Laboratory, Corporate Research and Development Center,
Toshiba Corporation, 1, Komukai Toshiba-cho, Saiwai-ku, Kawasaki 212-8582, Japan.
}

\date{Received; accepted; published online}

\abstract{
We have studied the generation of magnetic fields by the Biermann mechanism
 in the pair-instability supernovae explosions of the first stars.
The Biermann
 mechanism produces magnetic fields in the shocked region between the
 bubble and interstellar medium (ISM), even if magnetic fields are
 absent initially. We have performed a series of two-dimensional
 magnetohydrodynamic simulations with the Biermann term and estimate the
 amplitude and total energy of the produced magnetic fields. We find
 that magnetic fields with amplitude $10^{-14}-10^{-17}$ G are generated
 inside the bubble, though the amount of magnetic fields generated
 depend on specific values of initial conditions. This corresponds to
 magnetic fields with total energy of $10^{28}-10^{31}$ erg per each supernova remnant,
 which is strong enough to be the seed magnetic field for a galactic
 and/or interstellar dynamo.   
}

\correspondence{email:hanayama@th.nao.ac.jp}

\maketitle

\section{Introduction}

Magnetic fields are ubiquitous in the universe. In fact, observations of
rotation measure and synchrotron radiation have revealed that magnetic
fields exist in astronomical objects with various scales: galaxies,
clusters of galaxies, extra-cluster fields, etc. (for a review, see,
e.g., Widrow 2002). The observed galactic magnetic fields have
both coherent and fluctuating components whose strengths are comparable
to each other (Fosalba et al. 2002; Han et al. 2004). 
Conventionally, these magnetic fields
are considered to be amplified and maintained by a dynamo mechanism. The
coherent component in a galaxy is expected to be amplified by a galactic
dynamo, while the fluctuating component may be amplified by an interstellar
dynamo driven by turbulent motion of the interstellar medium (ISM; Balsara et al.2004). However, the dynamo mechanism itself cannot explain the
origin of the magnetic fields: seed magnetic fields are needed for 
the dynamo mechanism to work.

Thus far, various mechanisms have been suggested as a possible source of
the seed field. They can be classified into two types, those of astrophysical origin and those of cosmological origin. Here we concentrate on the former (for the latter scenario, see, e.g., Davies \& Widrow 2000; Takahashi et al. 2005; Amjad \& Robert 2005; Tsagas 2005; Yamazaki et al. 2005). 
Basically, the astrophysical magnetogenesis invokes the Biermann
mechanism (Biermann 1950), which is induced by the electric currents
produced when the spatial gradient of the electron pressure is not parallel to
that of the density. This is a pure plasma effect so that there is no
need to assume unknown physics as is often done in cosmological
models. Because the Biermann mechanism requires a non-parallel spatial
gradient of the pressure and density, some nonadiabatic process is
necessary to produce deviation from a polytropic equation of state. 
The strong magnetic fields in high-redshift galaxies 
(Athreya et al. 1998) imply that a significant amount of the seed magnetic
field should be generated at an early stage, e.g., the epoch of cosmological
reionization or protogalaxy formation. For instance, Gnedin et al.(2000)
studied the generation of magnetic fields in the ionizing front and
found that magnetic fields as high as $\approx 10^{-18}$ G in virialized
objects can be generated. Kulsrud et al.(1997) showed that a magnetic field of
$\approx 10^{-21}$ G can also be generated at shocks of large-scale
structure formation. 

In this paper, we investigate magnetogenesis at smaller
scales. Specifically, we study the amplitude of the magnetic field 
produced by the Biermann mechanism when the shock waves of the
supernova explosions of the first stars are spreading throughout 
the ISM (Hanayama et al. 2005).
The primordial supernova explosions are expected to take place
effectively, since the initial mass function (IMF) of Population III
stars should be substantially top-heavy (e.g., Abel et al. 2002).
We perform a series of two-dimensional magnetohydrodynamic (MHD)
numerical simulations in which the Biermann term is included.  We also
discuss whether they can be the origin of the cosmic magnetic
fields. Consequently, we find that the spatially averaged amplitude
of the produced magnetic field in virialized objects reaches $\sim 10^{-16}$ G,
which is much greater than those expected from cosmic reionization
and large-scale structure formation. Thus, the supernova explosions of the
first stars can be effective sources for the seed magnetic fields.
Although the situation considered here is somewhat similar to
that of Miranda et al.(1998), they assumed a multiple-explosion scenario of
structure formation and considered explosions of objects with mass
$>10^6M_\odot$ at $z\ge 100$, which is clearly unrealistic in the
context of the current standard model of structure formation.

\section{Numerical simulations and results}
We solve the two-dimensional MHD equations using conserved quantities with heating and cooling adjusted for the ISM. The induction equation with Biermann term is as follows:
\begin{equation}
\frac{\partial \mathbf{B}}{\partial t}
= \nabla \times (\mathbf{v} \times \mathbf{B})
  + \alpha \frac{\nabla \rho \times \nabla P}{\rho^{2}}
\label{ind}
\end{equation}
where $\rho$, $\mathbf{v}$, $P$, and $\mathbf{B}$ are the density,
velocity, pressure, and magnetic field, respectively. 
The last term of the
right-hand side of equation (\ref{ind}) is the Biermann term:  $\alpha$ 
in equation (\ref{ind}) is the so-called Biermann coupling constant
defined by $\alpha = m_{p} c/e (1 + \chi) \sim 10^{-4}~{\rm G~s}$,
where $m_{p}$, $e$, and $\chi$ are the proton mass, electric charge,
and ionization fraction, respectively. Although the gas temperature just
before star formation begins is rather cool ($T \sim 200 {\rm K}$; Abel et al. 2002; Bromm et al. 2003; Omukai \& Palla 2003), UV radiation from the first stars ionizes the
surrounding ISM (Freyer et al. 2003; Mori et al. 2004). Thus we set $\chi=5/6$ assuming 
$n_{He}/n_{H} \sim 0.1$. We use a cooling function derived
by Raymond et al.(1976). When parcels cool below $10^{4}$K, an artificial
heating rate proportional to the density is used. The constant heating coefficient is set so that heating balances cooling at the ambient density and temperature. Although the
cooling function and heating rate in the primordial gas are not clear so
far, they are not important for the adiabatic expansion phase that we
concentrate on. 

We solve the above equations by the two-dimensional MHD code in 
cylindrical coordinates $(r, z, \phi)$ assuming axial symmetry around
the symmetry axis ($z$). The code is based on the modified Lax-Wendroff
scheme with an artificial viscosity of von Neumann and Richtmyer to
capture shocks. The numerical scheme was tested by comparing known
solutions that have been obtained either analytically and numerically (see e.g., Hanayama \& Tomisaka 2005).  

In all the computations, grid spacings are chosen $\Delta r=\Delta z=0.1$ pc.
For example, the numerical domain covers a region of 130 pc $\times$ 130 pc 
with 1300 ($r$) $\times$ 1300 ($z$) mesh points in our fiducial model. 
We begin the simulation by adding a thermal energy of $E_0=10^{53}$ or $ 
10^{52}~ {\rm erg}$ within the sphere of 2~pc in radius.   

As the bubble expands, the ejected gas interacts with the ISM, and a shock wave
is formed. In the shocked region the gas is heated nonadiabatically,
which is a necessary condition for the Biermann mechanism to work. Here
the structure of the interstellar environment is important because it
affects the density and pressure profiles of the shocked region that is
directly related to the Biermann term. We assume an inhomogeneous  ISM with
average density $n_{\rm ISM} = 0.2~ {\rm cm}^{-3}$. This is roughly
consistent with the situation discussed in Bromm et al.(2003). The scale
length of the density and the amplitude of the inhomogeneity are as yet poorly
understood, and we assume inhomogeneity with the scale length
$\lambda = 1 {\rm ~pc}$ and density variation $0.2 \times 2^{\pm 1} {\rm cm}^{-3}$, which are values similar to those in our galaxy. 
Within the variation, the amplitude of the density is given at random 
and the distribution is smoothed numerically to create a perturbation. 
This is our fiducial model for the ISM. As we show in the next section, 
the amplitude of  the produced magnetic field is sensitive to the scale length
$\lambda$, while the average density and the amplitude of the density
variation are rather unimportant. Thus we consider several different
models in addition to the fiducial model: specifically, we vary the mean
density [$1 \times 2^{\pm 1} {\rm ~cm}^{-3}$ and $10 \times 2^{\pm 1} {\rm ~cm}^{-3}$] and the scale length ($3$ and $10~ {\rm pc}$). 

As for the explosion energy of the supernova, we adopt $E_{\rm SN} = 10^{53} {\rm ~erg}$ for the fiducial mode. This explosion energy corresponds to stars with mass $250 M_{\odot}$ that explode as a
pair-instability supernova (Fryer et al. 2001). In addition, we consider a model with
$E_{\rm SN} = 10^{52} {\rm ~erg}$ as a variation. 

\begin{figure}
\resizebox{\hsize}{!}
{\includegraphics[width=25cm,clip]{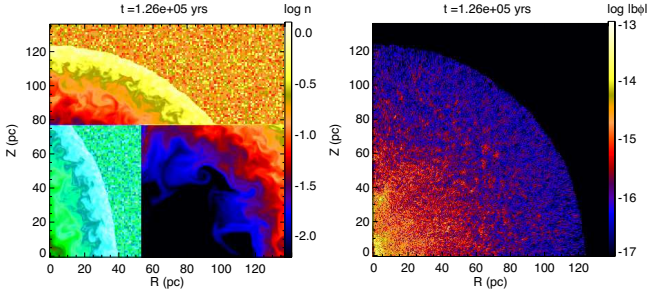}}
\caption{
Contours of the gas density ({\it left}) and the magnetic field ({\it right}) at the end of
the adiabatic expansion phase ($t = 1.26 \times 10^{5} {\rm yr}$) for the fiducial model.
The amplitude of the magnetic field is about $10^{-14} {\rm ~G}$ at the central region
and about $10^{-17} {\rm ~G}$ just behind the shock.
}
\label{fig:contour}
\end{figure}

Figure \ref{fig:contour} shows the contours of the gas density and the
amplitude of the magnetic field produced by the end of the adiabatic
expansion phase $t = 1.26 \times 10^{5} \rm{~yr}$ for the fiducial
model. The radius of the bubble is about 125 pc, and turbulent motion is
induced in the shocked region because of the inhomogeneity of the ISM. The
amplitude of the magnetic field is about $10^{-14} {\rm ~G}$ for the
central region and about $10^{-17} {\rm ~G}$ just behind the shock. 
The total magnetic energy inside the bubble is about $10^{30} {\rm ~erg}$. 

We have also checked the case of a homogeneous medium to test the robustness of
our computation. We find that, on average, a SNR generates magnetic fields with
 an amplitude $\sim10^{-19}$ G behind the shock front, which is
$\sim10^{-3}$ times smaller than that in an inhomogeneous medium. Therefore,
in the case of a homogeneous medium, we estimate a numerical error of the
amplitude of the magnetic field of $\sim10^{-19}$ G.  It should be noted
that the resulting magnetic field has only a toroidal component because
axial symmetry was assumed.  

\begin{figure}
\resizebox{\hsize}{!}
{\includegraphics[width=1cm,clip]{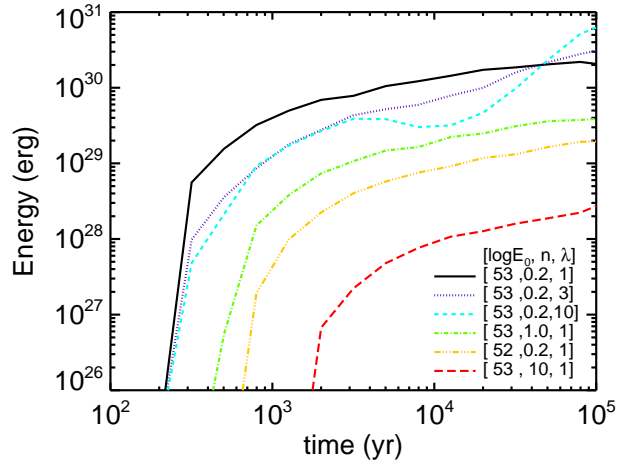}}
\caption{Time evolution of the total magnetic energy produced by the
 Biermann mechanism for various models.
\label{fig:evolution}}
\end{figure}

In Figure \ref{fig:evolution}, we show the time evolution of the total
magnetic energy for various models. The behaviors are qualitatively
similar for all the models. The robust knees around $10^3$ yr in
Figure \ref{fig:evolution} come from the formation of an adiabatic shock
front: this corresponds to when the SNR shifts from the free expansion phase to
the Sedov phase. The total magnetic energy at the end of the adiabatic
expansion phase  is larger for models with a smaller scale length of the
ISM density.  This tendency is confirmed by an order-of-magnitude
estimate in the next section. For models with large average ISM density
or small explosion energy,  the total magnetic energy is smaller because
the bubble is smaller than in the other models. Although there are many
uncertainties in the initial conditions,  the generation of a magnetic field
with a total energy of $10^{28}-10^{31} {\rm ~erg}$ appears to be robust.

\section{Analytic estimates of magnetic fields and implications for the seed magnetic Field}

To understand the result of the numerical simulations, in this section we
perform an order-of-magnitude estimation of the strength of the magnetic
field produced by the Biermann mechanism. 
The amplitude of the magnetic field produced by the Biermann mechanism can
be estimated from the Biermann term in equation (\ref{ind}),
\begin{equation}
B_{\rm Biermann} \sim \alpha \frac{\nabla \rho \times \nabla P}{\rho^{2}} \Delta t,
\end{equation}
where $\Delta t \sim 10^{3}~{\rm yr}$ is the characteristic timescale in which 
the Biermann mechanism works. Taking the characteristic pressure to be the
ram pressure of the gas, $P \sim P_{\rm ram} = \rho~v_{\rm bub}^{2}$,
and the characteristic velocity  of the bubble $v_{\rm bub} \sim
10^{-3}~{\rm pc\,yr^{-1}}$, we obtain
\begin{eqnarray}
B_{\rm Biermann}
&\sim& \alpha \frac{v_{\rm bub}^{2}}{\lambda L} \Delta t \nonumber \\
&\sim& 3 \times 10^{-15}
       \left(\frac{v_{\rm bub}}{10^{-3} {\rm pc\,yr^{-1}}}\right)^{2}
       \left(\frac{\lambda}{1{\rm pc}}\right)^{-1} \nonumber \\
& &    \times \left(\frac{L}{1{\rm pc}}\right)^{-1}
       \left(\frac{\Delta t}{10^{3}{\rm yr}}\right)~{\rm G},
\end{eqnarray}
where $L$ is the scale length of the pressure component perpendicular to
the density gradient. Then the magnetic energy produced by the Biermann
mechanism for each primordial SNR can be estimated as
\begin{eqnarray}
E_{B} &\sim& 
\frac{4 \pi R_{\rm bub}^{3}}{3} \frac{B_{\rm Biermann}^{2}}{8 \pi} 
\nonumber \\
&\sim& 5 \times 10^{31}
       \left(\frac{R_{\rm bub}}{100{\rm pc}}\right)^{3}
       \left(\frac{v_{\rm bub}}{10^{-3} {\rm pc\,yr^{-1}}}\right)^{4}
       \left(\frac{\lambda}{1{\rm pc}}\right)^{-2} \nonumber \\
& &    \times \left(\frac{L}{1{\rm pc}}\right)^{-2}
       \left(\frac{\Delta t}{10^{3}{\rm yr}}\right)^2 {\rm erg},\label{eb}
\label{eq:E-estimate}
\end{eqnarray}
which is consistent with the value obtained from our numerical simulations.

The dependence of the total magnetic energy on several parameters can 
also be understood from equation (\ref{eb}). It is found directly from equation (\ref{eb}) that $E_B\propto
\lambda^{-2}$. To examine the
dependence on the other parameters, we simply assume the Sedov-Taylor
solution: 
\begin{equation}
R_{\rm bub} \propto t^{2/5}\left(\frac{E_{\rm SN}}{n_{\rm ISM}}\right)^{1/5}, 
v_{\rm bub} \propto t^{-3/5}\left(\frac{E_{\rm SN}}{n_{\rm ISM}}\right)^{1/5}.
\end{equation}
Putting these into equation (\ref{eb}) yields 
$E_B\propto(E_{\rm SN}/n_{\rm ISM})^{7/5}$. Our numerical results shown
in Figure \ref{fig:evolution} are quite consistent with these simple
estimations. 

Now we estimate the spatially averaged energy density of the magnetic
fields produced by the first stars and consider whether they can be a source of
the seed fields. For the primordial star formation rate, we
extrapolate the one by Pell{\' o} et al.(2004) and Ricotti et al.(2004);  
$\dot{\rho_{\star}} \sim 10^{-2}~M_{\odot}~{\rm yr}^{-1} {\rm Mpc}^{-3}$.
Denoting the magnetic energy produced by the Biermann mechanism as
$\epsilon_{SN} \sim 10^{30} {\rm erg}$, the magnetic energy density
produced during the formation period of the first-star ($\tau \sim 1~{\rm Gyr}$) can be obtained as 
\begin{eqnarray}
e_{\rm B} & \sim & f_{\gamma\gamma}\dot{\rho_{\star}}~
\Bigl(\frac{\epsilon_{SN}}{M_{\rm SN}}\Bigr)\tau \nonumber \\
&\sim& 10^{-40}
       \left(\frac{f_{\gamma\gamma}}{0.06}\right)
       \left(\frac{\dot{\rho_{\star}}}{10^{-2}~M_{\odot}~{\rm yr}^{-1}}\right)
       \left(\frac{M_{\rm SN}}{250 M_{\odot}} \right)^{-1} \nonumber \\
& &    \times \left(\frac{\epsilon_{SN}}{10^{30}{\rm erg}}\right)
       \left(\frac{\tau}{1~{\rm Gyr}}\right) {\rm erg\,cm^{-3}},
\label{eq:10}
\end{eqnarray}
where $M_{\rm SN}$ is the typical mass scale of first
stars that end up in pair-instability supernovae, and $f_{\gamma\gamma}$
is the mass fraction of such first stars; we adopt
$f_{\gamma\gamma}=0.06$, which was derived under the assumption that very
massive black holes produced from the first stars end up in supermassive
black holes in galactic centers (Schneider et al. 2002). We note that the value is the
comoving density averaged in the universe; we can convert the value to
physical density in virialized objects (i.e., protogalaxies) as
\begin{eqnarray}
 e_{\rm B,gal} &\sim& e_{\rm B}(1+z)^4\Delta \sim 10^{-34}
\left(\frac{e_{\rm B}}{10^{-40}{\rm erg\,cm^{-3}}}\right) \nonumber \\
& & \times \left(\frac{1+z}{10}\right)^4
\left(\frac{\Delta}{200}\right) {\rm erg\,cm^{-3}},
\label{eq:11}
\end{eqnarray}
where $\Delta$ is the density contrast. This corresponds to the mean
magnetic field of $B\sim 10^{-16}$ G in protogalaxies, which is much
stronger than that expected in ionizing fronts, $B \sim 10^{-18}$ G
 (Gnedin et al. 2000). 

\section{Summary and discussion}
\label{sec:4}

We have studied the generation of magnetic fields in primordial
supernova remnants. We have performed two-dimensional MHD simulations
with the Biermann term, which can produce a magnetic field through
the nonadiabatic interaction between the bubble and ISM, even if there
is no magnetic field at first. We have found that the ISM around the
primordial supernovae is an effective site for producing magnetic fields. 
The total energy of the magnetic fields is $10^{28}-10^{31} {\rm
~erg}$, depending on the parameters adopted. On the basis of the results, 
we have estimated the spatially averaged energy density of the magnetic fields
produced by the first stars during the formation period of the first
stars. The averaged energy density is about $10^{-40} {\rm ~erg~cm}^{-3}$, 
which corresponds to $B\sim 10^{-16}$ G in protogalaxies at
$z\sim 10$.  This is much greater than expected from cosmic reionization
and large-scale structure formation. Thus primordial supernova remnants would
be a promising source for the seed fields for the galactic and/or
interstellar dynamo.  

Although the coherence length of the seed field computed here is much
smaller than the galactic scale, it can be amplified by the galactic
dynamo to produce a coherent component if the coherence length is
about 100 pc (Poezd et al. 1993; Ferri{\' e}re \& Schmitt 2000), which is a typical size of
supernova remnants. It might also be amplified by an interstellar dynamo to
produce the fluctuating component (Balsara et al. 2004). While it is beyond our
scope to discuss the relation between the magnetic field produced and the 
galactic/interstellar dynamo, we plan to investigate the evolutions of
the seed magnetic fields computed here as a result of the dynamo
processes on the large scale of galaxies and clusters of galaxies. This
will be presented in a forthcoming paper (H. Hanayama et al. 2005, in
preparation).  

Regarding their use as an observational signature, a proposal to detect 
seed magnetic fields was made by, e.g., Plaga(1995). If there exist 
intergalactic magnetic fields produced by the primordial SNRs, the arrival 
time of high energy gamma-ray photons from extragalactic sources would be 
delayed by the action of intergalactic magnetic fields on electron cascades
 (Ando 2004). 
Even a magnetic field as weak as $\sim10^{-24}$ G would be detectable if the
delay of the arrival time comes within a reasonable range (a few days).  
Therefore the mechanism of magnetic field generation proposed in this paper 
might be tested by the future high-energy gamma-ray experiments such as 
{\it GLAST} 
({\it Gamma-Ray Large Area Space Telescope}). 

\acknowledgements
We would like to thank Kohji Tomisaka for invaluable comments and advice 
on writing the paper. Numerical computations were carried out on
the VPP5000 supercomputer  at the Astronomical Data Analysis Center of
the National Astronomical Observatory, Japan. We also thank Ryoji
Matsumoto and Takaaki Yokoyama for a contribution to the calculation
code, CANS(Coordinated Astronomical Numerical Software).  K. T.,
M. O., and K. I. are supported by a Grant-in-Aid for JSPS Fellows.



\begin{thebibliography}{}
\bibitem{} Abel, T., Bryan, G.~L., Norman, M.~L.: 2002, Science 295, 93 

\bibitem{} Amjad, A., Robert B.~M.: 2005, Phys.Rev. D 71, 103509

\bibitem{} Ando, S.: 2004, MNRAS 354, 414 

\bibitem{} Athreya, R.~M., Kapahi,
 V.~K., McCarthy, P.~J., van Breugel, W.: 1998, A\&A 329, 809 

\bibitem{} Balsara, D.~S., Kim, J., Mac Low, M., Mathews, G.~J.: 2004, ApJ 617, 339 

\bibitem{} Biermann L.: 1950, Z. Naturforsch 5a, 65

\bibitem{} Bromm, V., Yoshida, N.,  Hernquist, L.: 2003, ApJ 596, L135

\bibitem{} Davies, G.,  Widrow, L.~M.: 2000, ApJ 540, 755

\bibitem{} Ferri\`{e}re, K., Schmitt, D.: 2000, A\&A 358, 125

\bibitem{} Fosalba, P., Lazarian, A., Prunet, S.,  Tauber, J.~A.: 2002, ApJ 564, 762

\bibitem{} Freyer, T., Hensler, G.,  Yorke, H.~W.: 2003, ApJ 594, 888

\bibitem{} Fryer, C.~L., Woosley, S.~E.,  Heger, A.: 2001, ApJ 550, 372

\bibitem{} Gnedin, N.~Y., Ferrara, A.,  Zweibel, E.~G.: 2000, ApJ 539, 505 

\bibitem{} Han, J.~L., Ferriere, K.,  Manchester, R.~N.: 2004, ApJ 610, 820 

\bibitem{} Hanayama, H., Tomisaka, K.: 2005, ApJ in press, astro-ph/0507421

\bibitem{} Hanayama, H., Takahashi, K., Kotake, K., Oguri, M., Ichiki, K., Ohno, H.: 2005, ApJ 633, 941

\bibitem{} Kulsrud, R.~M., Cen, R., Ostriker, J.~P., Ryu, D.: 1997, ApJ 480, 481
\bibitem{} Miranda, O.~D., Opher, M., Opher, R.: 1998, MNRAS 301, 547 

\bibitem{} Mori, M., Umemura, M., Ferrara, A.: 2004, ApJ 613, L97 

\bibitem{} Omukai, K.,  Palla, F.: 2003, ApJ 589, 677 

\bibitem{} Pell{\' o}, R., Schaerer, D., Richard, J., Le Borgne, J.-F., Kneib, J.-P.: 2004, A\&A 416, L35 

\bibitem{} Plaga, R.: 1995, Nature 374, 430 

\bibitem{} Poezd, A., Shukurov, D., Sokoloff, D.: 1993, MNRAS 264, 285

\bibitem{} Raymond, J.~C., Cox, D.~P.,  Smith, B.~W.: 1976, ApJ 204, 290 

\bibitem{} Ricotti, M., Haehnelt, M.~G., Pettini, M.,  Rees, M.~J.: 2004, MNRAS 352, L21 

\bibitem{} Schneider, R., Ferrara, A., Natarajan, P., Omukai, K.: 2002, ApJ 571, 30 

\bibitem{} Tsagas, C.~G.: 2005, Phys.Rev. D 72, 123509 

\bibitem{} Takahashi, K., Ichiki, K., Ohno, H., Hanayama, H.: 2005, Phys. Rev. Lett. 95, 121301

\bibitem{} Widrow, L.~M.: 2002, Rev. Mod. Phys. 74, 775 

\bibitem{} Yamazaki, D.~G., Ichiki, K., Kajino, T.: 2005, ApJ 625, L1 
\end{thebibliography}
\end{document}